\documentclass[reprint,superscriptaddress,pra]{revtex4-2}

\usepackage{graphicx}
\usepackage{dcolumn}
\usepackage{bm}
\usepackage{graphicx}
\usepackage{bm}
\usepackage{pgfplots}
\usepackage{color}
\usepackage{algorithm}
\usepackage{url,array}
\usepackage{amsmath,amssymb,amsthm,dsfont,mathrsfs}
\usepackage{epstopdf}

\def\x{{\mathbf{x}}}
\def\y{{\mathbf{y}}}

\def\M{{\mathbf{M}}}
\def\A{{\mathbf{A}}}
\def\I{{\mathbf{I}}}
\def\s{{\mathbf{s}}}

\def\c{{\mathbf{c}}}

\def\u{{\mathbf{u}}}

\def\H{{\mathbf{H}}}
\def\v{{\mathbf{v}}}

\def\n{{\mathbf{n}}}

\begin{document}


\title{Joint parameter estimation and multidimensional reconciliation for continuous-variable quantum key distribution}

\author{Jisheng Dai} \email{jsdai@dhu.edu.cn}
\affiliation{College of Information Science and Technology, Donghua University, Shanghai 201620, China}
\author{Xue-Qin Jiang} \email{xqjiang@dhu.edu.cn}
\affiliation{College of Information Science and Technology, Donghua University, Shanghai 201620, China}
\affiliation{Hefei National Laboratory, Hefei 230088, China}

\author{Peng Huang}
\author{Tao Wang}
\author{Guihua Zeng}
\affiliation{State Key Laboratory of Advanced Optical Communication Systems and Networks, Shanghai Jiao Tong University, Shanghai 200240, China}
\affiliation{Hefei National Laboratory, Hefei 230088, China}


\begin{abstract}
Accurate quantum channel parameter estimation is essential for effective information reconciliation in continuous-variable quantum key distribution (CV-QKD).
However, conventional maximum likelihood (ML) estimators rely on a large amount of disclosed data, leading to a significant loss in symbol efficiency. Moreover, the separation between the estimation and reconciliation phases can introduce error propagation.
In this paper, we propose a novel joint message-passing scheme that unifies channel parameter estimation and information reconciliation within a Bayesian framework. By leveraging the expectation-maximization (EM) algorithm, the proposed method simultaneously estimates unknown parameters during decoding, eliminating the need for separate ML estimation.
Furthermore, we introduce a hybrid multidimensional rotation scheme that removes the requirement for norm feedback, significantly reducing classical channel overhead. To the best of our knowledge, this is the first work to unify multidimensional reconciliation and channel parameter estimation in CV-QKD, providing a practical solution for high-efficiency reconciliation with minimal information disclosure.
\end{abstract}

\maketitle


\section{Introduction}

In the modern digital era, the importance of information security continues to grow. Quantum key distribution (QKD) allows two remote users, typically referred to as Alice and Bob, to generate an absolutely secure cryptographic key through the transmission of quantum signals. It remains secure even if an unauthorized party attempts to intercept the communication \cite{BEN84,RevModPhys.74.145}.
A specific implementation, continuous-variable (CV) QKD \cite{PhysRevLett.88.057902,PhysRevA.76.042305,Zhang2024APR}, encodes information using the quadrature measurements of quantum states. This approach offers strong compatibility with conventional coherent optical communication infrastructure \cite{PhysRevA.95.062330,Bian2024APL,Zhang_2019,Wang_2024}. Due to this advantage, CV-QKD can achieve significantly improved key generation rates, especially in medium-range transmissions. Additionally, it leverages mature modulation and demodulation techniques, enhancing overall efficiency \cite{Yang2023EPJ}.
For CV-QKD systems employing Gaussian-modulated coherent states, rigorous security analyses have confirmed their resilience against diverse attack strategies \cite{PhysRevLett.101.200504,PhysRevLett.114.070501,PhysRevLett.118.200501}.

A CV-QKD system operates in two main stages: quantum transmission and post-processing \cite{Laudenbach2018}. During quantum transmission, Alice sends encoded quantum states to Bob through a quantum channel, where Bob performs measurements using homodyne or heterodyne detection techniques \cite{PhysRevA.76.042305}. The subsequent post-processing phase derives secure cryptographic keys through multiple steps, including basis sifting, parameter estimation, error correction (information reconciliation), and privacy amplification \cite{RevModPhys.74.145,Yang2023EPJ}.
Among these steps, information reconciliation plays a critical role, as it corrects discrepancies in the raw key data using error-correcting codes. Its efficiency significantly influences both the secret key rate (SKR) and the maximum achievable transmission distance \cite{PhysRevApplied.8.044017,PhysRevLett.125.010502}. Due to its direct impact on system performance, optimizing information reconciliation remains a major research focus in advancing CV-QKD technology \cite{Laudenbach2018,ZhouC2019,Jiang2024}.

CV-QKD systems employ multiple reconciliation strategies, with slice reconciliation \cite{1266817} and multidimensional reconciliation \cite{PhysRevA.77.042325} being the most prominent.
Slice reconciliation operates through quantization and error correction, whereas multidimensional reconciliation incorporates a transformation step before error correction.
A key advantage of multidimensional reconciliation is its use of a $D$-dimensional rotation to convert the physical Gaussian channel into a virtual binary-input additive white Gaussian noise channel (BI-AWGNC) \cite{WOS:000432546600001}. When integrated with advanced error-correcting codes (e.g., low-density parity-check (LDPC) \cite{PhysRevA.84.062317,Zhangkun2025} or polar codes \cite{Liu_2025}), this method substantially improves reconciliation efficiency and secret key rates, particularly under low signal-to-noise ratio (SNR) conditions.
Extensive research has been devoted to enhancing multidimensional reconciliation.
For example, norm-free decoding was introduced in \cite{Li2019QIP}, eliminating the need for encoder-supplied norm data.
A Student's t-distribution noise model was adopted in \cite{PhysRevA.103.032603} for improved decoding.
Non-Gaussian modulation protocols were adapted in \cite{PhysRevApplied.19.054084}, further boosting CV-QKD performance.
Lossy compression-based polar codes for high-throughput CV-QKD were proposed in \cite{Liu_2025}.
Recently, a cross-rotation scheme was proposed to overcome the restriction that closed-form rotations are limited to at most eight dimensions, enabling reconciliation in arbitrarily high-dimensional spaces \cite{q3s4-ht52}.
Deep learning-based multidimensional reconciliation schemes were addressed in \cite{Feng_2025,photonics9020110}.
All these advancements share a common objective: optimizing reconciliation efficiency through refined encoding and decoding techniques.

On the other hand, accurate quantum channel parameter estimation plays a crucial role in the performance of multidimensional reconciliation \cite{Matsuura2021,PhysRevA.81.062343}. The conventional maximum likelihood (ML) estimator requires a large amount of disclosed data to achieve reliable estimates.
For example, the number of disclosed symbols is set to half of the available sequences in \cite{PhysRevA.81.062343}. Consequently, a substantial portion of the symbols must be revealed for parameter estimation, reducing the effective symbol efficiency for information reconciliation.
Moreover, in the traditional ML-based approach, parameter estimation and information reconciliation are treated as two distinct phases, leading to possible error propagation from the estimation phase into the reconciliation phase.
Therefore, it is essential to explore alternative methods for estimating quantum channel parameters both accurately and efficiently.
We observe that the disclosed symbols and reconciliation symbols experience the same channel fading, and by leveraging the shared channel parameters, it is possible to significantly improve estimation performance while reducing the amount of disclosed data required for parameter estimation. Specifically, the number of disclosed symbols required to achieve a given estimation accuracy can be notably reduced.
To the best of our knowledge, however, no existing work has addressed a joint design that integrates error correction for information reconciliation with channel parameter estimation.

In this paper, we propose a new joint message-passing scheme that simultaneously estimates channel parameters and performs information reconciliation. The key idea is to cast the combined problem of parameter estimation and multidimensional reconciliation into a unified Bayesian framework and use the expectation-maximization (EM) algorithm to automatically learn the unknown parameters.
For efficient Bayesian inference, we further construct a factor graph that incorporates the LDPC code's factor graph as a subgraph and develop a generalized sum-product message-passing algorithm for joint LDPC decoding and parameter estimation.
By eliminating the need for a separate ML estimator, our approach avoids its associated performance degradation and significantly enhances both parameter estimation accuracy and multidimensional reconciliation efficiency.
Moreover, we introduce a novel hybrid multidimensional rotation scheme that circumvents the need for norm feedback, thereby substantially reducing the feedback overhead on the classical channel.

\section{Review of ML estimator and its limitations}

In this section, we first review the well-known ML estimator for quantum channel parameter estimation, and then discuss its limitations regarding data efficiency and error propagation.

\subsection{Parameter estimation with ML estimator}

In a CV-QKD system, after quantum state transmission and data sifting, Alice and Bob obtain correlated Gaussian sequences of length $L$, denoted by $\x \in \mathbb{R}^{L}$ and $\y \in \mathbb{R}^{L}$, respectively.
For ease of presentation, we partition these sequences into two subsequences:
$\x = [(\x^p)^{\mathrm{T}}, (\x^d)^{\mathrm{T}} ]^{\mathrm{T}}$ and $\y = [(\y^p)^{\mathrm{T}}, (\y^d)^{\mathrm{T}} ]^{\mathrm{T}}$,
where $\x^p \in \mathbb{R}^{M}$ and $\y^p \in \mathbb{R}^{M}$ are disclosed for quantum channel parameter estimation,
while $\x^d \in \mathbb{R}^{N}$ and $\y^d \in \mathbb{R}^{N}$ are used to distill the secret key between the two parties.
Here, $(\cdot)^{\mathrm{T}}$ denotes vector/matrix transpose, and $L = M + N$.
The paired $\{ \x^p, \y^p \}$ and $\{ \x^d, \y^d \}$ can be modeled as \cite{PhysRevA.81.062343}:
\begin{align}
\y^p=&  t \cdot \x^p  +  \n^p ,\label{eq-yxz1}
\end{align}
and
\begin{align}
\y^d=&  t \cdot \x^d  +  \n^d ,\label{eq-yxz2}
\end{align}
where $t\triangleq\sqrt{\eta T}$, $\eta$ denotes the homodyne detector efficiency, $T$ denotes the channel transmittance, $\n^p \in \mathbb{R}^{M} $ (or $\n^d \in \mathbb{R}^{N} $) denotes the Gaussian quantum channel noise vector with zero mean and variance $\sigma^2\triangleq 1+v_\mathrm{el}+ \eta T\varepsilon $, $v_\mathrm{el}$ denotes the added electronic noise variance in shot-noise units,
and $\varepsilon$ denotes the excess channel noise variance in shot-noise units.

In order to distill secret keys between Alice and Bob, the channel parameters $t$ and $\sigma^2$ must be accurately estimated in advance.
Using the disclosed symbols $\x^p$ and $\y^p$, the well-known ML estimator is given by \cite{PhysRevA.81.062343}:
\begin{align}
\hat{t} &= \frac{(\x^p)^{\mathrm{T}}\y^p}{\|\x^p\|^2}, \label{eq-LS1} \\
\hat{\sigma}^2 &= \frac{\|\y^p - \hat{t} \cdot \x^p\|^2}{M}, \label{eq-LS2}
\end{align}
where $\|\cdot\|$ denotes the vector norm.
Then, by substituting these estimated parameters into (\ref{eq-yxz2}), information reconciliation is performed to correct errors and establish shared secret keys between the two parties.
{To simplify the presentation, in the following we use the term \lq\lq pilot symbols'' to denote the disclosed symbols $\x^p$ and $\y^p$ used for parameter estimation, rather than optical reference signals for phase recovery or local-oscillator synchronization.
}

\subsection{Limitations of ML estimator}

As shown in \cite{PhysRevA.81.062343}, the parameter estimation accuracy has a significant impact on the SKR.
The ML estimator mentioned in \cite{PhysRevA.81.062343} requires a large number of pilot symbols to achieve reliable estimation. As a result, a substantial fraction of transmitted symbols must be disclosed for parameter estimation, which significantly reduces the number of symbols available for information reconciliation.
Another limitation is that the ML estimator treats parameter estimation and information reconciliation as separate processes, which can lead to error propagation from the estimation phase into the reconciliation process.
Although such error propagation due to separate operations has been extensively studied in the wireless communications community (e.g., \cite{10111055,9716058}), it has received relatively little attention in the context of CV-QKD.

Our key observation is that the paired subsequences $\{ \x^d, \y^d \}$ used for information reconciliation experience the same channel fading as the pilots.
By jointly performing parameter estimation and information reconciliation using the complete set of paired subsequences $\{ \x^p, \y^p \}$ and $\{ \x^d, \y^d \}$, we can enhance both parameter estimation accuracy and information reconciliation performance.
Equivalently, this approach allows for a significant reduction in the number of pilots required to achieve a target estimation accuracy.
In the following, we present a new joint message-passing scheme to realize this goal.
The core innovation lies in the formulation of a unified Bayesian framework for joint parameter estimation and multidimensional reconciliation, where optimal parameter estimates are derived in the Bayesian sense using the EM algorithm \cite{543975}.

While Bayesian parameter estimation methods have been proposed in \cite{PhysRevA.99.032326,PhysRevApplied.18.054077}, they are designed for different purposes and rely on prior knowledge of the quantum channel.
Moreover, these methods treat parameter estimation and information reconciliation as separate processes, so the issue of error propagation due to decoupled operation still persists.
In contrast, our proposed approach jointly addresses both tasks and is particularly well-suited for free-space CV-QKD scenarios with time-varying channels \cite{Jing:22}, as it requires only a small number of pilots to accurately estimate the channel parameters.

\section{Proposed method}
In this section, we first introduce an improved hybrid multidimensional rotation scheme that eliminates the need for norm feedback.
Next, we present a unified Bayesian framework for joint parameter estimation and information reconciliation, along with the corresponding factor graph.
We then derive a generalized sum-product message-passing procedure for efficient Bayesian inference.
Finally, we demonstrate how the unknown parameters are learned using the EM algorithm.


\subsection{Improved hybrid multidimensional rotation}

To maintain a stable SKR over long transmission distances, {multidimensional reconciliation \cite{PhysRevA.77.042325}} is adopted in the following procedure.
On Bob's side, he divides the received data sequence $\y^d$ into blocks:  $\y^d=[(\y_1^d)^{\mathrm{T}},(\y_2^d)^{\mathrm{T}},\ldots, (\y_G^d)^{\mathrm{T}} ]^{\mathrm{T}}$,
where $\y_g^d \in \mathbb{R}^D $, $D\in\{1,2,4,8\}$, and $G=N/D$.
He then selects a parity-check matrix {$\H \in \mathbb{F}_2^{L\times N}$} and generates a codeword
$\c=[\c_1^{\mathrm{T}}, \c_2^{\mathrm{T}},\ldots, \c_G^{\mathrm{T}} ]^{\mathrm{T}}$ with {$\c_g \in\mathbb{F}_2^D $}.
{Note that the sequence \(\c\) is not necessarily a valid codeword satisfying \(\H\c=\mathbf{0}\). Instead, Bob transmits the syndrome $\s = \H\c$ to Alice for syndrome-based LDPC decoding.}
For each bit string $\c_g$, it is transformed into a $D$-dimensional spherical vector:
\begin{align}
\u_g=\left[ \frac{ (-1)^{c_{1,g}}}{\sqrt{D}} , \frac{ (-1)^{c_{2,g}}}{\sqrt{D}},\ldots, \frac{ (-1)^{c_{D,g}}}{\sqrt{D}}     \right]^{\mathrm{T}}. \label{eq-ug1D}
\end{align}
If $D = 1, 2, 4$, or $8$, there always exists a unitary matrix $\M_g$ that establishes the mapping between $\y_g^d$ and $\u_g$ \cite{PhysRevA.77.042325}:
\begin{align}
\underbrace{\sum_{i=1}^D \alpha_{i,g} \A_i}_{\triangleq \M_g} \frac{\y_g^d}{\|\y_g^d\|} = \u_g, \label{eq-mygug}
\end{align}
where $\alpha_{i,g} \triangleq \u_g^{\mathrm{T}} \A_i \y_g^d / \|\y_g^d\|$ is the $i$-th mapping coefficient and $\A_i \in \mathbb{R}^{D \times D}$ denotes a fixed orthogonal matrix (defined in the Appendix of \cite{PhysRevA.77.042325}).
Note that the recently proposed cross-rotation scheme~\cite{q3s4-ht52} can extend the reconciliation dimension to arbitrarily high values. However, this topic is beyond the scope of this paper and is therefore not discussed here.
{As far as the multidimensional rotation scheme \cite{PhysRevA.77.042325} itself is concerned, it does not require the transmission of the vector norm \(\|\y_g^d\|\).
However, when multidimensional rotation is combined with soft-decision LDPC decoding \cite{PhysRevA.84.062317}, the vector norm \(\|\y_g^d\|\) becomes necessary.}
Therefore, to assist Alice in secret-key distillation, Bob usually needs to transmit the mapping coefficients $\alpha_{i,g}$, the vector norms $\|\y_g^d\|$, and the syndrome $\s$ over the classical channel.

We argue that the vector norms and mapping coefficients do not need to be sent to Alice \emph{independently}. Instead, we can transmit a set of hybrid combined coefficients that encapsulate the same rotation information as the original scheme, thereby significantly reducing the feedback load over the classical channel. Specifically, let (\ref{eq-mygug}) be rewritten as:
\begin{align}\label{eq-mygug-beta}
\underbrace{\sum_{i=1}^D \beta_{i,g} \A_i}_{= \M_g / \| \y_g^d \|} \y_g^d = \u_g,
\end{align}
where $\beta_{i,g} \triangleq \alpha_{i,g} / \| \y_g^d \|$. Clearly, the hybrid mapping coefficients $\beta_{i,g}$s contain sufficient information to reconstruct $\M_g / \| \y_g^d \|$.
Therefore, in the proposed hybrid rotation scheme, Bob only needs to send the hybrid coefficients $\beta_{i,g}$ and the syndrome $\s$ to Alice over the classical channel.

On Alice's side, she divides the sent data sequence $\x^d$ into $\x^d = [(\x_1^d)^{\mathrm{T}}, (\x_2^d)^{\mathrm{T}}, \ldots, (\x_G^d)^{\mathrm{T}} ]^{\mathrm{T}}$ with $\x_g^d \in \mathbb{R}^D$. Then, (\ref{eq-yxz2}) can be rewritten as:
\begin{align}
\x^d_g = \frac{1}{t}\left( \y^d_g - \n^d_g \right). \label{eq-txyn}
\end{align}
Once Alice receives the hybrid mapping coefficients $\beta_{i,g}$, she can directly construct $\M_g / \| \y_g^d \|$ without first reconstructing $\M_g$ and then dividing by $\| \y_g^d \|$. {It is worth noting that \(\|\y_g^d\|\) can be perfectly recovered according to
\begin{align}
\Big(\frac{\M_g}{\| \y_g^d \|}\Big)^{\mathrm{T}}
\frac{\M_g}{\| \y_g^d \|}
=
\frac{\M_g^{\mathrm{T}}\M_g}{\| \y_g^d \|^2}
=
\frac{\I}{\| \y_g^d \|^2},
\end{align}
where the last equality follows from the fact that \(\M_g\) is a unitary matrix.}
Left-multiplying both sides of (\ref{eq-txyn}) by $\M_g / \| \y_g^d \|$ yields:
\begin{align}
\underbrace{\frac{\M_g}{\| \y_g^d \|} \x^d_g}_{\triangleq \v_g}
= \frac{1}{t} \left( \frac{\M_g}{\| \y_g^d \|} \y^d_g - \frac{\M_g}{\| \y_g^d \|} \n^d_g \right)
= \frac{1}{t} \left( \u_g + \frac{\bar{\n}_g}{\| \y_g^d \|} \right),  \label{eq-txyn3}
\end{align}
where the second equality follows from (\ref{eq-mygug}), and $\bar{\n}_g$ denotes the rotated noise vector, which retains the same Gaussian distribution as $\n^d_g$ since $\M_g$ is a unitary matrix \cite{Harpaz2005}.
Note that both the channel gain $t$ and the noise variance $\sigma^2$ remain unknown at this stage, as the proposed method does not rely on the ML estimators for parameter estimation.

The goal of the proposed scheme on Alice's side is to jointly estimate the channel parameters ($t$ and $\sigma^2$) and decode $\u$ (or equivalently, $\c$) using the sum-product message-passing algorithm \cite{910572}, leveraging the observations in (\ref{eq-yxz1}), (\ref{eq-txyn3}), and the syndrome $\s$. Decoding is considered successful if the estimated codeword $\hat{\c}$ satisfies the parity-check condition:
\begin{align}
\H \hat{\c} = \s,
\end{align}
where $\hat{\c}$ denotes the estimate of the transmitted codeword $\c$.

\subsection{Joint Bayesian formulation and its factor graph}

To jointly perform parameter estimation and LDPC decoding, we formulate the problem within a unified Bayesian framework and construct the corresponding factor graph \cite{1267047} in detail. This factor graph serves as the foundation for Bayesian inference via the sum-product message-passing algorithm. In the following, both the channel gain $t$ and the noise variance $\sigma^2$ are treated as deterministic parameters to be estimated, rather than random variables, and are iteratively updated using the EM algorithm.

Under the i.i.d. Gaussian noise assumption, the pilot model (\ref{eq-yxz1}) and the data model (\ref{eq-txyn3}) lead to the following probability density functions (PDFs):
\begin{align}
p(\y^p; t, \sigma^2)= \prod_{m=1}^M \mathcal{N} \left( y^p_m | t x^p_m,  \sigma^2  \right),
\end{align}
and
\begin{align}
p(\v|\u; t, \sigma^2)=&\prod_{g=1}^G p(\v_g|\u_g; t, \sigma^2)\notag\\
=& \prod_{g=1}^G \prod_{i=1}^D \mathcal{N} \Big( v_{i,g} \Big| \frac{u_{i,g}}{t}   ,  \frac{\sigma^2}{t^2\|\y_g^d\|^2}  \Big),
\end{align}
respectively, where $y^p_m$  ($x^p_m$) denotes the $m$-th element of $\y^p$ ($\x^p$), and $v_{i,g}$ ($u_{i,g}$) denotes the $i$-th element of $\v_g$ ($\u_g$). From (\ref{eq-ug1D}), the conditional PDF $p(\c|\u)$ can be expressed as
\begin{align}
p(\c|\u)= \prod_{g=1}^G p(\c_g|\u_g)= \prod_{g=1}^G \prod_{i=1}^D   p( c_{i,g} |u_{i,g}),
\end{align}
where
\begin{align}
 p( c_{i,g} |u_{i,g})=
 \begin{cases}
  \delta(c_{i,g}),  &u_{i,g}=\frac{1}{\sqrt{D}}\\
  \delta(c_{i,g}-1),  &u_{i,g}=-\frac{1}{\sqrt{D}}
 \end{cases},
\end{align}
and $\delta(\cdot)$ denotes the Dirac delta function.
Moreover, $p(\u)$ follows a discrete uniform distributions:
\begin{align}
p(\u)=\prod_{g=1}^G \prod_{i=1}^D  \mathrm{Uniform_D}(u_{i,g}),
\end{align}
where
\begin{align}
  \mathrm{Uniform_D}(u_{i,g})=  \frac{1}{2}\delta\Big(u_{i,g} -  \frac{1}{\sqrt{D}}  \Big) +  \frac{1}{2}\delta\Big(u_{i,g} +  \frac{1}{\sqrt{D}}  \Big) .
\end{align}
The syndrome constrain $\s=\H\c$ gives:
\begin{align}
p(\s|\c)= \delta(\s-\H\c).
\end{align}

\begin{figure*}
  \centering
  \includegraphics[scale=0.8]{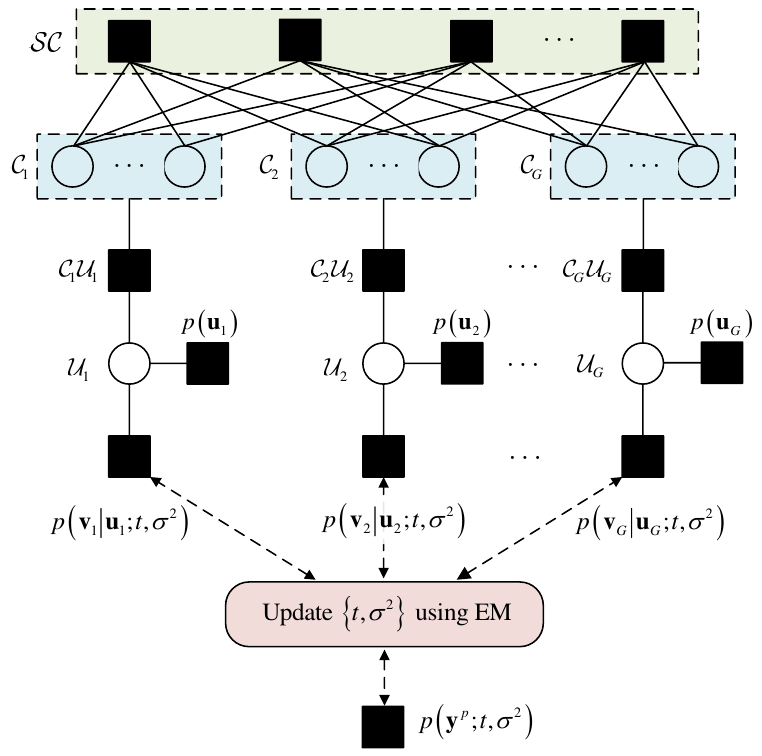}
    \caption{Illustration of the factor graph used for message passing in the proposed joint parameter estimation and information reconciliation scheme. Variable nodes are represented by blank circles, and factor nodes by black squares.}
\label{fg}
\end{figure*}

Therefore, the joint PDF can be expressed as:
\begin{align}
&p(\y^p, \v, \u, \c; t, \sigma^2)\notag\\
=& p(\y^p; t, \sigma^2) \, p(\v|\u; t, \sigma^2) \, p(\c|\u) \, p(\u) \, p(\s|\c).
\end{align}
The corresponding factor graph is shown in Fig.~\ref{fg}, where blank circles and black squares represent variable nodes and factor nodes, respectively. Note that the variable nodes $\mathcal{C}_g$ and $\mathcal{U}_g$ correspond to $\c_g$ and $\u_g$, respectively, and the distributions associated with the factor nodes $\mathcal{C}_g\mathcal{U}_g$ and $\mathcal{SC}$ are given by $p(\c_g|\u_g)$ and $p(\s|\c)$, respectively.

\subsection{Generalized sum-product message passing}

Let \(\Delta_{a \rightarrow b}(\cdot)\) denote the message passed from Node \(a\) to Node \(b\). According to the sum-product update rule \cite{910572}, the message passed from Node \(\mathcal{U}_g\) to Node \(\mathcal{C}_g \mathcal{U}_g\) can be written as:
\begin{align}
&\Delta_{\mathcal{U}_g \rightarrow \mathcal{C}_g\mathcal{U}_g}(\u_g)\notag\\
\propto& p(\u_g)  p(\v_g|\u_g; t, \sigma^2)\notag\\
\propto&  \prod_{i=1}^D  \Bigg\{     \delta\Big(u_{i,g} -  \frac{1}{\sqrt{D}}  \Big)       \mathcal{N} \Big( v_{i,g} \Big| \frac{1}{t\sqrt{D}}   ,  \frac{\sigma^2}{t^2\|\y_g^d\|^2}  \Big)\notag\\
&~~~~~~~+
\delta\Big(u_{i,g} +  \frac{1}{\sqrt{D}}  \Big)       \mathcal{N} \Big( v_{i,g} \Big| \frac{-1}{t\sqrt{D}}   ,  \frac{\sigma^2}{t^2\|\y_g^d\|^2}  \Big)
\Bigg\}.
\end{align}
The message passed from Node $\mathcal{C}_g\mathcal{U}_g$ to Node $\mathcal{C}_g$ can be written as:
\begin{align}
&\Delta_{\mathcal{C}_g\mathcal{U}_g \rightarrow \mathcal{C}_g}(\c_g) \notag\\
\propto&
\sum_{\u_g}  p(\c_g| \u_g) \Delta_{\mathcal{U}_g \rightarrow \mathcal{C}_g\mathcal{U}_g}(\u_g)\notag\\
\propto&  \prod_{i=1}^D  \Bigg\{     \delta(c_{i,g} )      \mathcal{N} \Big( v_{i,g} \Big| \frac{1}{t\sqrt{D}}   ,  \frac{\sigma^2}{t^2\|\y_g^d\|^2}  \Big)\notag\\
&~~~~~~~+
\delta(c_{i,g} -1 )        \mathcal{N} \Big( v_{i,g} \Big| \frac{-1}{t\sqrt{D}}   ,  \frac{\sigma^2}{t^2\|\y_g^d\|^2}  \Big)
\Bigg\}.\label{cucg}
\end{align}
Using (\ref{cucg}),  the initial log-likelihood ratio (LLR) for the LDPC decoding can be calculated by:
\begin{align}
\mathrm{LLR}_{i,g}=  \ln \frac{ \mathcal{N} \Big( v_{i,g} \Big| \frac{1}{t\sqrt{D}}   ,  \frac{\sigma^2}{t^2\|\y_g^d\|^2}  \Big)}{\mathcal{N} \Big( v_{i,g} \Big| \frac{-1}{t\sqrt{D}}   ,  \frac{\sigma^2}{t^2\|\y_g^d\|^2}  \Big)}
=\frac{ 2 t v_{i,g} \|\y_g^d\|^2  }{\sqrt{D}\sigma^2}.\label{eq-LLRig}
\end{align}
It is worth noting that although $t$ and $\sigma^2$ appear in~(\ref{eq-LLRig}) and subsequent expressions, their true values are not required, as they are automatically learned through the iterative EM updates (see Section~III-D).
Inputting the LLR into the LDPC decoder, we are able to update the message passed from Node $\mathcal{SC}$ to Node $\mathcal{C}_g$, which is denoted as:
\begin{align}
\Delta_{\mathcal{SC}\rightarrow \mathcal{C}_g}(\c_g)
\propto\prod_{i=1}^D  \Big\{    p_{i,g}^{0} \delta(c_{i,g} ) +
 p_{i,g}^{1} \delta(c_{i,g} -1 )
\Big\},\label{cucg2}
\end{align}
where \( p_{i,g}^{0} \) and \( p_{i,g}^{1} \) denote the probabilities output by the LDPC decoder that \( c_{i,g}=0 \) and \( c_{i,g}=1 \), respectively.
Then, the message passed from Node $\mathcal{C}_g\mathcal{U}_g$ to Node $\mathcal{U}_g$ becomes:
\begin{align}
&\Delta_{\mathcal{C}_g\mathcal{U}_g \rightarrow \mathcal{U}_g}(\u_g) \notag\\
\propto&
\prod_{i=1}^D  \Big\{  p_{i,g}^{0} \delta\Big(u_{i,g} -  \frac{1}{\sqrt{D}}  \Big) +  p_{i,g}^{1}\delta\Big(u_{i,g} +  \frac{1}{\sqrt{D}}  \Big)
\Big\}.\label{cucg3}
\end{align}
The sum-product belief at Node $\mathcal{U}_g$ can be calculated by:
\begin{align}
&b_{sp}(\u_g) \notag\\
\propto&   \Delta_{\mathcal{C}_g\mathcal{U}_g \rightarrow \mathcal{U}_g}(\u_g)  p(\u_g)   p(\v|\u ; t, \sigma^2 )\notag\\
\propto&  \prod_{i=1}^D  \Bigg\{    p_{i,g}^{0}   \delta\Big(u_{i,g} -  \frac{1}{\sqrt{D}}  \Big)       \mathcal{N} \Big( v_{i,g} \Big| \frac{1}{t\sqrt{D}}   ,  \frac{\sigma^2}{t^2\|\y_g^d\|^2}  \Big)\notag\\
&~~~~~~~+
 p_{i,g}^{1}  \delta\Big(u_{i,g} +  \frac{1}{\sqrt{D}}  \Big)       \mathcal{N} \Big( v_{i,g} \Big| \frac{-1}{t\sqrt{D}}   ,  \frac{\sigma^2}{t^2\|\y_g^d\|^2}  \Big)
\Bigg\}.
\end{align}
Therefore, the posterior distribution $p(\u|\v; t,\sigma^2)$ is given as:
\begin{align}
&p(\u|\v; t,\sigma^2)\notag\\
=& \prod_{g}^G p(\u_g|\v_g; t,\sigma^2)  \notag\\
= &  \prod_{g}^G \prod_{i=1}^D  \Bigg\{    {p}_{i,g}^{+}   \delta\Big(u_{i,g} -  \frac{1}{\sqrt{D}}  \Big)  +
 {p}_{i,g}^{-}  \delta\Big(u_{i,g} +  \frac{1}{\sqrt{D}}  \Big)
\Bigg\},\label{eq-puvtsigma}
\end{align}
where
\begin{align}
 p_{i,g}^{+}=&  \frac{ p_{i,g}^{0}}{p_{i,g}^{0}  +  p_{i,g}^{1} e^{  - \frac{ 2 t v_{i,g} \|\y_g^d\|^2  }{\sqrt{D}\sigma^2}  }      },\\
 p_{i,g}^{-}=&  \frac{ p_{i,g}^{1}}{p_{i,g}^{1}  +  p_{i,g}^{0} e^{   \frac{ 2 t v_{i,g} \|\y_g^d\|^2  }{\sqrt{D}\sigma^2}  }      }.
\end{align}

\subsection{EM-driven parameter learning}

Since the parameters $t$ and $\sigma^2$ are usually unknown in practice, the EM algorithm will be
utilized to tune these parameters automatically. $\u$ and $\c$
are taken as the hidden variables, and the unknown parameters are updated as \cite{9716058,543975,6898015}:
\begin{align}
\{t, \sigma^2\} =& \arg\max_{t, \sigma^2}  \mathbb{E}\Big\{  \ln   p(\y^p, \v, \u, \c; t, \sigma^2 )     \Big\}  \notag\\
=&  \arg\max_{t, \sigma^2}  \ln  p(\y^p; t, \sigma^2) + \mathbb{E}\Big\{  \ln p(\v|\u ; t, \sigma^2 )       \Big\}  \notag\\
=&  \arg\max_{t, \sigma^2} -\frac{L}{2}\ln \sigma^2    - \frac{\| \y^p - t\x^p   \|^2}{2\sigma^2}\notag\\
 &~~~~~~~~~~~~~~~~~-  \mathbb{E}\Bigg\{  \sum_{g=1}^G  \frac{\|\y_g^d\|^2}{2\sigma^2}  \| t\v_g- \u_g  \|^2 \Bigg\}\notag\\
=&  \arg\max_{t, \sigma^2} -\frac{L}{2}\ln \sigma^2    - \frac{\| \y^p - t\x^p   \|^2}{2\sigma^2}\notag\\
&~-  \sum_{g=1}^G\frac{\|\y_g^d\|^2}{2\sigma^2} \Big( \| t\v_g- \mathbb{E}\{\u_g\}  \|^2  + \mathrm{Tr}\big( \mathbb{D}\{\u_g \}\big)  \Big),\label{eq-EMobject}
\end{align}
where  $\mathbb{E}\{\cdot\}$ and $\mathbb{D}\{\cdot\}$ denote the expectation and  variance with respect to $p(\u|\v; t,\sigma^2)$, respectively, $\mathrm{Tr}(\cdot)$ denotes the trace operation, $\mathbb{E}\{u_{i,g}\}$ is calculated by $( p_{i,g}^{+} - p_{i,g}^{-}  )/\sqrt{D}$, and $\mathbb{D}\{u_{i,g} \}$ is calculated by
$ \mathbb{E}\{u_{i,g}^2\} - (\mathbb{E}\{u_{i,g}\})^2 =  4 p_{i,g}^{+} p_{i,g}^{-}/D$.

Taking the derivative of the objective function in (\ref{eq-EMobject}) with respect to $t$ and setting it to zero yields:
\begin{align}
\frac{  (\x^p)^{\mathrm{T}}( \y^p - t\x^p  )}{2\sigma^2}    +   \sum_{g=1}^G \frac{\|\y_g^d\|^2 \v_g^{\mathrm{T}}( \mathbb{E}\{\u_g\} -t\v_g )}{2\sigma^2} =0.
\end{align}
Therefore, the optimal update of $t$ is:
\begin{align}
t^{\mathrm{new}}=  \frac{(\x^p)^{\mathrm{T}} \y^p +  \sum_{g=1}^G \|\y_g^d\|^2 \v_g^{\mathrm{T}} \mathbb{E}\{\u_g\}   }
{(\x^p)^{\mathrm{T}}\x^p  +  \sum_{g=1}^G \|\y_g^d\|^2 \v_g^{\mathrm{T}}\v_g }.\label{eq-tnew}
\end{align}
Taking the derivative of the objective function in (\ref{eq-EMobject}) with respect to $\sigma^2$ and setting it to zero yields:
\begin{align}
&\sum_{g=1}^G\frac{\|\y_g^d\|^2}{2(\sigma^2)^{2}} \Big( \| t\v_g- \mathbb{E}\{\u_g\}  \|^2  + \mathrm{Tr}\big( \mathbb{D}\{\u_g \}\big)  \Big)\notag\\
&~~~~~~~~~~~~~~~~~~~~~~~~~~~~~~-\frac{L}{2\sigma^2 } +  \frac{\| \y^p - t\x^p   \|^2}{2(\sigma^2)^{2}}  =  0.
\end{align}
Therefore, the optimal update of $\sigma^2$ is:
\begin{align}
(\sigma^2)^{\mathrm{new}} =&  \frac{\| \y^p - t\x^p   \|^2}{L} \notag\\
&+   \sum_{g=1}^G \frac{\|\y_g^d\|^2}{L} \Big( \| t\v_g- \mathbb{E}\{\u_g\}  \|^2  + \mathrm{Tr}\big( \mathbb{D}\{\u_g \}\big)  \Big).\label{sigmanew}
\end{align}

The proposed message passing algorithm proceeds by repeating (\ref{eq-LLRig}), (\ref{cucg2}),
(\ref{eq-puvtsigma}), (\ref{eq-tnew}) and (\ref{sigmanew}), respectively.
Note that we need the initializations of \(t\) and \(\sigma^2\) to trigger the proposed message passing algorithm. These parameters can be simply initialized using estimators based on the pilot signals.

\subsection{Performance and complexity analysis}
In this subsection, we begin with proving that the proposed joint parameter estimation and multidimensional reconciliation scheme does not lead to any information leakage.
Firstly, the pilots $\mathbf{y}^p$ and $\mathbf{x}^p$ do not contain any secret key information. Therefore, disclosing these pilots does not result in information leakage. Secondly, since the proposed hybrid multidimensional rotation follows the same orthogonal transformation as the conventional multidimensional reconciliation, it can be shown that the mapping coefficients $\{\mathbf{M}_g,\ \forall g\}$ are statistically independent of $\mathbf{u}$, by following the procedure in Section~V-C of~\cite{PhysRevA.77.042325} or using the sum-product message-passing technique in Section~III-E of \cite{q3s4-ht52}. Consequently, the proposed hybrid multidimensional rotation does not reveal any information about the secret keys. Thirdly, the proposed EM-based scheme involves only signal processing operations and does not require any additional prior information; hence, it also does not cause information leakage.

{Then, we analyze the computational complexity of the proposed method and show that it remains in the same order as that of conventional multidimensional reconciliation. Recall that the proposed EM-based scheme iteratively performs the updates in (\ref{eq-LLRig}), (\ref{cucg2}), (\ref{eq-puvtsigma}), (\ref{eq-tnew}), and (\ref{sigmanew}). The main computational complexity in each iteration is summarized as follows.
\begin{itemize}
\item The computational complexity of calculating every \(\mathrm{LLR}_{i,g}\) is \(\mathcal{O}(D)\). Since (\ref{eq-LLRig}) needs to be evaluated \(G\) times in each iteration, its total computational complexity is \(\mathcal{O}(N)\) per iteration.
\item The total computational complexity of (\ref{cucg2}) is dominated by the soft-decision LDPC decoding in the log domain, which is \(\mathcal{O}(N d_v d_c)\) per iteration, where \(d_v\) and \(d_c\) denote the average variable-node degree and average check-node degree, respectively.
\item The total computational complexity of (\ref{eq-puvtsigma}) is \(\mathcal{O}(N)\) per iteration, since both \(p_{i,g}^{+}\) and \(p_{i,g}^{-}\) only involve element-wise multiplications and additions.
\item It is easy to verify that the computational complexities of (\ref{eq-tnew}) and (\ref{sigmanew}) are both \(\mathcal{O}(N)\) per iteration.
\end{itemize}
Since conventional multidimensional reconciliation already includes the operations in (\ref{eq-LLRig}) and (\ref{cucg2}), its overall computational complexity is \(\mathcal{O}(N d_v d_c I_{\mathrm{iter}})\), where \(I_{\mathrm{iter}}\) denotes the number of LDPC decoding iterations. Clearly, this complexity is dominated by the LDPC decoding process in (\ref{cucg2}).
For the proposed method, the additional operations in (\ref{eq-puvtsigma}), (\ref{eq-tnew}), and (\ref{sigmanew}) only introduce \(\mathcal{O}(N)\) computational overhead, which is negligible compared with the complexity of LDPC decoding. Therefore, the overall computational complexity of the proposed method is also \(\mathcal{O}(N d_v d_c I_{\mathrm{iter}})\), remaining in the same order as that of conventional multidimensional reconciliation.
}

{Finally, we intuitively explain that the proposed EM-based joint estimation framework can reduce the number of disclosed pilot symbols while maintaining the same parameter-estimation security level \(\epsilon_{PE}\) \cite{PhysRevA.81.062343}. In conventional CV-QKD systems, parameter estimation is performed solely based on the disclosed pilot symbols. Therefore, the estimation accuracy and the corresponding security parameter \(\epsilon_{PE}\) are directly determined by the number of disclosed samples. In contrast, in the proposed method, the LDPC-based information reconciliation accompanied by joint parameter estimation enables the recovered reconciliation symbols to be reliably obtained during the decoding process and further exploited for parameter estimation. Consequently, the proposed scheme effectively performs parameter estimation using both the disclosed pilot symbols and the recovered reconciliation symbols.
From the perspective of estimation theory, this is equivalent to increasing the effective sample size used for parameter estimation. Therefore, compared with conventional pilot-only estimation, the proposed method can achieve the same estimation accuracy and maintain the same \(\epsilon_{PE}\) with a substantially smaller number of disclosed pilot symbols.}

\section{Simulation and performance evaluation}

This section provides several simulation tests to demonstrate the advantages of the proposed method.
For one-way CV-QKD reverse reconciliation, we utilize an ATSC~3.0 LDPC code~\cite{e22101087} with a codeword length of \(N = 64800\) bits, and the code rate is set to \(R = 0.2\) or \(R =0.1333\).
The proposed EM-driven message-passing technique is evaluated against the conventional ML-based approach \cite{PhysRevA.81.062343} and {the maximum \emph{a posteriori} (MAP)-based approach \cite{PhysRevApplied.18.054077}} for multidimensional reverse reconciliation, considering dimensions of
\( D = 4 \) and \( 8 \).
{We assume that the MAP-based approach has perfect knowledge of the channel and noise prior distributions, whereas neither the proposed method nor the ML-based approach relies on such prior information.}
All LDPC decoding strategies employ sum-product decoding with a maximum iteration limit of 200.
For a given SNR $\kappa$, the reconciliation efficiency is calculated as:
\begin{align}\label{eq-bs}
\beta(\kappa)= \frac{R}{0.5\log_2(1 + \kappa)}.
\end{align}
Conversely, for a given reconciliation efficiency \(\beta\), {the corresponding SNR is calculated as:
\begin{align}
\kappa(\beta) = 2^{2R/\beta} - 1. \label{eq-kappa}
\end{align}}

In Simulation~1, we conduct Monte Carlo trials to evaluate the parameter estimation performance, using the root mean square error (RMSE) as the performance metric. The RMSEs for the channel gain and noise variance estimates are defined as:
\begin{align}
\mathrm{RMSE}_t = \sqrt{\frac{1}{J} \sum_{j=1}^{J} (\hat{t}_{(j)} - t_{(j)})^2},
\end{align}
and
\begin{align}
\mathrm{RMSE}_{\sigma^2} = \sqrt{\frac{1}{J} \sum_{j=1}^{J} (\hat{\sigma}^2_{(j)} - \sigma_{(j)}^2)^2},
\end{align}
respectively,
where \( J=5000 \) is the total number of Monte Carlo trials, \( \hat{t}_{(j)} \) is the estimate of the channel gain \( t_{(j)} \) at the \( j \)-th trial, and
\( \hat{\sigma}^2_{(j)} \) is the estimate of the noise variance \( \sigma^2_{(j)} \) at the \( j \)-th trial.
Fig.~\ref{figS1} shows the RMSE performance of two strategies (the proposed method and the traditional ML-based method)
versus the SNR, with the number of available pilots fixed at \( M=600 \) and the reconciliation dimension fixed at \( D=8 \).
The results show that, for both methods, the RMSE decreases as the SNR increases (or as the noise level decreases). Notably, the proposed EM-driven method significantly outperforms the traditional ML-based method, as it exploits the shared channel parameters between pilot and data symbols, whereas the traditional method relies solely on pilots for parameter estimation.

\begin{figure}
\center
\begin{tikzpicture}[scale=0.9]
\begin{semilogyaxis}[xlabel={SNR [dB]},title={a)},
ylabel={$\mathrm{RMSE}_t$},grid=major,
legend style={at={(0.2,0.225),font=\footnotesize}, legend cell align={left},
anchor=north,legend columns=1},  xmin=-4,xmax=12,
]
\addplot[mark=square]  coordinates{
  (    -4,   0.037770508249346  )
  (    -2,   0.030002181142342  )
  (     0,   0.023831579584674  )
  (     2,   0.018930096542186  )
  (     4,   0.015036710169515  )
  (     6,   0.011944083445011  )
  (     8,   0.009487522718275  )
  (    10,   0.007536207172714  )
  (    12,   0.005986222140018  )
};
\addplot[mark=asterisk,red]  coordinates{
  (    -4, 0.003583930611025 )
  (    -2, 0.002846807925041 )
  (     0, 0.002261299430253 )
  (     2, 0.001796213819873 )
  (     4, 0.001426783287742 )
  (     6, 0.001133334228269 )
  (     8, 0.000900239382271 )
  (    10, 0.000715085621860 )
  (    12, 0.000568012750614 )
};
\legend{ML, Proposed}
\end{semilogyaxis}
\end{tikzpicture}
\begin{tikzpicture}[scale=0.9]
\begin{semilogyaxis}[xlabel={SNR [dB]},title={b)},
ylabel={$\mathrm{RMSE}_{\sigma^2}$},grid=major,
legend style={at={(0.26,0.985),font=\footnotesize}, legend cell align={left},
anchor=north,legend columns=1}, xmin=-4,xmax=12
]
\addplot[mark=square]  coordinates{
  (    -4, 0.066049519084653 )
  (    -2, 0.041674429165846 )
  (     0, 0.026294787159210 )
  (     2, 0.016590889079647 )
  (     4, 0.010468143316260 )
  (     6, 0.006604951908465 )
  (     8, 0.004167442916585 )
  (    10, 0.002629478715921 )
  (    12, 0.001659088907965 )
};
\addplot[mark=asterisk,red]  coordinates{
  (    -4, 0.006226545940625 )
  (    -2, 0.003928683449271 )
  (     0, 0.002478831799507 )
  (     2, 0.001564037141977 )
  (     4, 0.000986840732785 )
  (     6, 0.000622654407935 )
  (     8, 0.000392868362208 )
  (    10, 0.000247883131591 )
  (    12, 0.000156403633610 )
};
\end{semilogyaxis}
\end{tikzpicture}
\caption{RMSE of parameter estimation versus SNR for an ATSC 3.0 LDPC code with rate $R=0.2$. a) RMSE for estimating $t$; and b) RMSE for estimating $\sigma^2$.
 }\label{figS1}
\end{figure}

In Simulation~2, we evaluate the bit error rate (BER) and frame error rate (FER) performance of the proposed method, the traditional ML-based method, and the MAP-based method at low SNRs (or high reconciliation efficiencies). The BER and FER are averaged over \(5000\) frames.
Figs.~\ref{figS2} and~\ref{figS222} present the BER and FER performance as functions of the SNR and the reconciliation efficiency under soft-decision LDPC decoding, where the code rate is set to \(R = 0.2\) and \(R = 0.1333\), respectively.
Note that SNR and reconciliation efficiency \(\beta\) correspond one-to-one as defined in (\ref{eq-bs}) and (\ref{eq-kappa}).
For reference, we also include an exhaustive ML method that uses
\({M=64800}\) pilots for channel parameter estimation, rather than \(M=600\) pilots.
The results reveal that:
1) the BER (or FER) of both methods improves as SNR increases (or as reconciliation efficiency decreases);
2) the proposed method consistently outperforms the traditional ML-based approach and the MAP-based approach at any fixed SNR (or \(\beta\)), regardless of the dimensionality of reconciliation; and
3) the proposed method achieves nearly the same BER and FER performance as the exhaustive ML method, demonstrating that it can significantly reduce the pilot overhead while maintaining excellent performance.

\begin{figure}
\center
\begin{tikzpicture}[scale=0.9]
\begin{semilogyaxis}[xlabel={SNR [dB]},title={a)},
ylabel={BER},grid=major,
legend style={at={(0.29,0.61),font=\footnotesize}, legend cell align={left},
anchor=north,legend columns=1},  xmin=-4.8546,xmax=-4.1378,ymax=0.36,ymin=0.0
]
\addplot[mark=triangle]  coordinates{
  (-4.1378, 0.003999074074074)
  (-4.1973, 0.009567901234567)
  (-4.2560, 0.022853024691358)
  (-4.3139, 0.052466496913580)
  (-4.3710, 0.101539259259259)
  (-4.4275, 0.161818462962963)
  (-4.4832, 0.211839956790124)
  (-4.5382, 0.243437240740741)
  (-4.5925, 0.257894746913580)
  (-4.6462, 0.264938953703704)
  (-4.6992, 0.269248555555555)
  (-4.7516, 0.272316006172840)
  (-4.8034, 0.275043929012346)
  (-4.8546, 0.277453824074074)
};
\addplot[mark=triangle*]  coordinates{
  (-4.1378, 0.001398996913580)
  (-4.1973, 0.003813811728395)
  (-4.2560, 0.012913577160494)
  (-4.3139, 0.037930833333333)
  (-4.3710, 0.084817185185185)
  (-4.4275, 0.147198932098765)
  (-4.4832, 0.204620123456790)
  (-4.5382, 0.238659913580247)
  (-4.5925, 0.255862080246914)
  (-4.6462, 0.263591512345679)
  (-4.6992, 0.268283549382716)
  (-4.7516, 0.271662888888889)
  (-4.8034, 0.274434524691358)
  (-4.8546, 0.276921336419753)
};
\addplot[mark=otimes,blue]  coordinates{
  (-4.1378, 0.000147620370370)
  (-4.1973, 0.000845293209876)
  (-4.2560, 0.005886395061728)
  (-4.3139, 0.023400506172839)
  (-4.3710, 0.063677410493827)
  (-4.4275, 0.126787543209877)
  (-4.4832, 0.188364043209877)
  (-4.5382, 0.230053706790123)
  (-4.5925, 0.251302512345679)
  (-4.6462, 0.261177496913580)
  (-4.6992, 0.266582314814815)
  (-4.7516, 0.270205132716049)
  (-4.8034, 0.273209867283951)
  (-4.8546, 0.275853111111111)
};
\addplot[mark=+,red]  coordinates{
  (-4.1378, 0.000148086419753)
  (-4.1973, 0.000918484567901)
  (-4.2560, 0.006116873456790)
  (-4.3139, 0.024354685185185)
  (-4.3710, 0.065648413580246)
  (-4.4275, 0.129169719135802)
  (-4.4832, 0.190487935185185)
  (-4.5382, 0.231461663580247)
  (-4.5925, 0.251726185185185)
  (-4.6462, 0.261373231481482)
  (-4.6992, 0.266710953703704)
  (-4.7516, 0.270330682098766)
  (-4.8034, 0.273315179012346)
  (-4.8546, 0.275944845679012)
};
\addplot[mark=square]  coordinates{
  (-4.1378, 0.000446913580246)
  (-4.1973, 0.001252583333333)
  (-4.2560, 0.001937663580246)
  (-4.3139, 0.005200787037037)
  (-4.3710, 0.010973962962963)
  (-4.4275, 0.025680972222222)
  (-4.4832, 0.057815620370370)
  (-4.5382, 0.105568037037037)
  (-4.5925, 0.161301108024691)
  (-4.6462, 0.209967293209877)
  (-4.6992, 0.239410734567902)
  (-4.7516, 0.254265419753086)
  (-4.8034, 0.261405756172840)
  (-4.8546, 0.265742521604938)
};
\addplot[mark=square*]  coordinates{
  (-4.1378,  0.000113561728395)
  (-4.1973,  0.000216040123457)
  (-4.2560,  0.000489672839506)
  (-4.3139,  0.001374376543210)
  (-4.3710,  0.004344635802469)
  (-4.4275,  0.015641876543210)
  (-4.4832,  0.042543422839506)
  (-4.5382,  0.091288882716049)
  (-4.5925,  0.150269398148148)
  (-4.6462,  0.201902746913580)
  (-4.6992,  0.235003953703704)
  (-4.7516,  0.251773660493827)
  (-4.8034,  0.260074910493827)
  (-4.8546,  0.264773228395062)
};
\addplot[mark=o,blue]  coordinates{
  (-4.1378, 0                )
  (-4.1973, 0                )
  (-4.2560, 0.000051151234567)
  (-4.3139, 0.000224222222222)
  (-4.3710, 0.001330833333333)
  (-4.4275, 0.007696635802469)
  (-4.4832, 0.028770410493827)
  (-4.5382, 0.070163882716049)
  (-4.5925, 0.128202404320988)
  (-4.6462, 0.186907586419753)
  (-4.6992, 0.226510966049383)
  (-4.7516, 0.247227030864198)
  (-4.8034, 0.257362246913580)
  (-4.8546, 0.262855895061729)
};
\addplot[mark=asterisk,red]  coordinates{
  (-4.1378, 0                )
  (-4.1973, 0                )
  (-4.2560, 0.000050870370370)
  (-4.3139, 0.000210543209876)
  (-4.3710, 0.001389629629629)
  (-4.4275, 0.008351774691358)
  (-4.4832, 0.029952552469135)
  (-4.5382, 0.071793172839506)
  (-4.5925, 0.130697719135802)
  (-4.6462, 0.189187157407407)
  (-4.6992, 0.227585811728395)
  (-4.7516, 0.247820049382716)
  (-4.8034, 0.257641354938272)
  (-4.8546, 0.263031148148148)
};
\legend{$D=4$ (ML), $D=4$ (MAP), $D=4$ (Exhaustive), $D=4$ (Proposed),$D=8$ (ML), $D=8$ (MAP), $D=8$ (Exhaustive), $D=8$ (Proposed)
}
\end{semilogyaxis}
\end{tikzpicture}
\begin{tikzpicture}[scale=0.9]
\begin{axis}[xlabel={$\beta$},title={b)},
ylabel={FER},grid=major,
legend style={at={(0.26,0.985),font=\footnotesize}, legend cell align={left},
anchor=north,legend columns=1}, xmin=0.85,xmax=0.98,ymax=1,ymin=0
]
\addplot[mark=triangle]  coordinates{
  (   0.8500, 0.0160000)
  (   0.8600, 0.0394000)
  (   0.8700, 0.0988000)
  (   0.8800, 0.2390000)
  (   0.8900, 0.4670000)
  (   0.9000, 0.7216000)
  (   0.9100, 0.8984000)
  (   0.9200, 0.9754000)
  (   0.9300, 0.9954000)
  (   0.9400, 0.9994000)
  (   0.9500, 1        )
  (   0.9600, 1        )
  (   0.9700, 1        )
  (   0.9800, 1        )
};
\addplot[mark=triangle*]  coordinates{
  (   0.8500,    0.00540000)
  (   0.8600,    0.01540000)
  (   0.8700,    0.05980000)
  (   0.8800,    0.18040000)
  (   0.8900,    0.40380000)
  (   0.9000,    0.67740000)
  (   0.9100,    0.88700000)
  (   0.9200,    0.97060000)
  (   0.9300,    0.99480000)
  (   0.9400,    0.99920000)
  (   0.9500,    1.00000000)
  (   0.9600,    1.00000000)
  (   0.9700,    1.00000000)
  (   0.9800,    1.00000000)
};
\addplot[mark=otimes,blue]  coordinates{
  (   0.8500, 0.00060000)
  (   0.8600, 0.00440000)
  (   0.8700, 0.03060000)
  (   0.8800, 0.12080000)
  (   0.8900, 0.32860000)
  (   0.9000, 0.61560000)
  (   0.9100, 0.84700000)
  (   0.9200, 0.95900000)
  (   0.9300, 0.99220000)
  (   0.9400, 0.99920000)
  (   0.9500, 0.99980000)
  (   0.9600, 1         )
  (   0.9700, 1         )
  (   0.9800, 1         )
};
\addplot[mark=+,red]  coordinates{
  (   0.8500, 0.00060000)
  (   0.8600, 0.00460000)
  (   0.8700, 0.03180000)
  (   0.8800, 0.12900000)
  (   0.8900, 0.34160000)
  (   0.9000, 0.62880000)
  (   0.9100, 0.85700000)
  (   0.9200, 0.96260000)
  (   0.9300, 0.99280000)
  (   0.9400, 0.99920000)
  (   0.9500, 0.99980000)
  (   0.9600, 1         )
  (   0.9700, 1         )
  (   0.9800, 1         )
};
\addplot[mark=square]  coordinates{
  (   0.8500, 0.0018000)
  (   0.8600, 0.0050000)
  (   0.8700, 0.0076000)
  (   0.8800, 0.0208000)
  (   0.8900, 0.0486000)
  (   0.9000, 0.1230000)
  (   0.9100, 0.2918000)
  (   0.9200, 0.5274000)
  (   0.9300, 0.7622000)
  (   0.9400, 0.9208000)
  (   0.9500, 0.9810000)
  (   0.9600, 0.9976000)
  (   0.9700, 1        )
  (   0.9800, 1        )
};
\addplot[mark=square*]  coordinates{
  (   0.8500,    0.00040000)
  (   0.8600,    0.00080000)
  (   0.8700,    0.00200000)
  (   0.8800,    0.00560000)
  (   0.8900,    0.02080000)
  (   0.9000,    0.08440000)
  (   0.9100,    0.22780000)
  (   0.9200,    0.47500000)
  (   0.9300,    0.73360000)
  (   0.9400,    0.90940000)
  (   0.9500,    0.97940000)
  (   0.9600,    0.99720000)
  (   0.9700,    0.99960000)
  (   0.9800,    0.99980000)
};
\addplot[mark=o,blue]  coordinates{
  (   0.8500, 0        )
  (   0.8600, 0        )
  (   0.8700, 0.0002000)
  (   0.8800, 0.0012000)
  (   0.8900, 0.0084000)
  (   0.9000, 0.0476000)
  (   0.9100, 0.1760000)
  (   0.9200, 0.4002000)
  (   0.9300, 0.6708000)
  (   0.9400, 0.8828000)
  (   0.9500, 0.9684000)
  (   0.9600, 0.9958000)
  (   0.9700, 1        )
  (   0.9800, 1        )
};
\addplot[mark=asterisk,red]  coordinates{
  (   0.8500, 0         )
  (   0.8600, 0         )
  (   0.8700, 0.00020000)
  (   0.8800, 0.00100000)
  (   0.8900, 0.00920000)
  (   0.9000, 0.05100000)
  (   0.9100, 0.18280000)
  (   0.9200, 0.41280000)
  (   0.9300, 0.68360000)
  (   0.9400, 0.89000000)
  (   0.9500, 0.97060000)
  (   0.9600, 0.99660000)
  (   0.9700, 1         )
  (   0.9800, 1         )
};
\end{axis}
\end{tikzpicture}
\caption{BER and FER for an ATSC 3.0 LDPC code with rate $R=0.2$. a) BER versus SNR; and b) FER versus $\beta$.
 }\label{figS2}
\end{figure}

\begin{figure}
\center
\begin{tikzpicture}[scale=0.9]
\begin{semilogyaxis}[xlabel={SNR [dB]},title={a)},
ylabel={BER},grid=major,
legend style={at={(0.29,0.61),font=\footnotesize}, legend cell align={left},
anchor=north,legend columns=1},  xmin=-6.8283,xmax=-6.1455,ymax=0.36,ymin=0.0
]
\addplot[mark=triangle]  coordinates{
  (-6.1455, 0.0079131)
  (-6.2020, 0.0133610)
  (-6.2577, 0.0217431)
  (-6.3128, 0.0374724)
  (-6.3672, 0.0640029)
  (-6.4209, 0.1053320)
  (-6.4739, 0.1563215)
  (-6.5264, 0.2089548)
  (-6.5782, 0.2518264)
  (-6.6293, 0.2811312)
  (-6.6799, 0.2980438)
  (-6.7300, 0.3080831)
  (-6.7794, 0.3137772)
  (-6.8283, 0.3174637)
};
\addplot[mark=triangle*]  coordinates{
  (-6.1455, 0.004788740740741)
  (-6.2020, 0.008396003086420)
  (-6.2577, 0.013786608024691)
  (-6.3128, 0.027932861111111)
  (-6.3672, 0.052268709876543)
  (-6.4209, 0.092696503086420)
  (-6.4739, 0.141796033950617)
  (-6.5264, 0.197422197530864)
  (-6.5782, 0.244547117283951)
  (-6.6293, 0.276756916666667)
  (-6.6799, 0.295712182098765)
  (-6.7300, 0.306502953703704)
  (-6.7794, 0.312927283950617)
  (-6.8283, 0.316826598765432)
};
\addplot[mark=otimes,blue]  coordinates{
  (-6.1455, 0.00018001)
  (-6.2020, 0.00073440)
  (-6.2577, 0.00241564)
  (-6.3128, 0.00759052)
  (-6.3672, 0.02320487)
  (-6.4209, 0.05499875)
  (-6.4739, 0.10227256)
  (-6.5264, 0.16150316)
  (-6.5782, 0.21627549)
  (-6.6293, 0.25785553)
  (-6.6799, 0.28346250)
  (-6.7300, 0.29871079)
  (-6.7794, 0.30759622)
  (-6.8283, 0.31316116)
};
\addplot[mark=+,red]  coordinates{
  (-6.1455, 0.00017831)
  (-6.2020, 0.00083211)
  (-6.2577, 0.00274863)
  (-6.3128, 0.00824558)
  (-6.3672, 0.02493692)
  (-6.4209, 0.05785379)
  (-6.4739, 0.10618061)
  (-6.5264, 0.16548807)
  (-6.5782, 0.22017948)
  (-6.6293, 0.26029987)
  (-6.6799, 0.28517130)
  (-6.7300, 0.29943718)
  (-6.7794, 0.30817894)
  (-6.8283, 0.31350940)
};
\addplot[mark=square]  coordinates{
  (-6.1455, 0.00295932)
  (-6.2020, 0.00495569)
  (-6.2577, 0.00749164)
  (-6.3128, 0.01228196)
  (-6.3672, 0.02005061)
  (-6.4209, 0.03569941)
  (-6.4739, 0.05912753)
  (-6.5264, 0.10026392)
  (-6.5782, 0.15010930)
  (-6.6293, 0.20121052)
  (-6.6799, 0.24430442)
  (-6.7300, 0.27387819)
  (-6.7794, 0.29225434)
  (-6.8283, 0.30339541)
};
\addplot[mark=square*]  coordinates{
  (-6.1455, 0.001443561728395)
  (-6.2020, 0.002218496913580)
  (-6.2577, 0.003619620370370)
  (-6.3128, 0.006538898148148)
  (-6.3672, 0.012516891975309)
  (-6.4209, 0.024759743827160)
  (-6.4739, 0.048214132716049)
  (-6.5264, 0.084934947530864)
  (-6.5782, 0.136364966049383)
  (-6.6293, 0.189937771604938)
  (-6.6799, 0.236003194444444)
  (-6.7300, 0.269799206790124)
  (-6.7794, 0.290583049382716)
  (-6.8283, 0.302325737654321)
};
\addplot[mark=o,blue]  coordinates{
  (-6.1455, 0)
  (-6.2020, 0)
  (-6.2577, 0.00015506)
  (-6.3128, 0.00046172)
  (-6.3672, 0.00174998)
  (-6.4209, 0.00719709)
  (-6.4739, 0.02115303)
  (-6.5264, 0.05113949)
  (-6.5782, 0.09733879)
  (-6.6293, 0.15444883)
  (-6.6799, 0.20771432)
  (-6.7300, 0.24827742)
  (-6.7794, 0.27620448)
  (-6.8283, 0.29272022)
};
\addplot[mark=asterisk,red]  coordinates{
  (-6.1455, 0)
  (-6.2020, 0)
  (-6.2577, 0.000174410)
  (-6.3128, 0.000635759)
  (-6.3672, 0.002284777)
  (-6.4209, 0.008621817)
  (-6.4739, 0.022898601)
  (-6.5264, 0.054526533)
  (-6.5782, 0.102578750)
  (-6.6293, 0.159438648)
  (-6.6799, 0.211909780)
  (-6.7300, 0.251287416)
  (-6.7794, 0.278281925)
  (-6.8283, 0.294066975)
};
\legend{$D=4$ (ML), $D=4$ (MAP), $D=4$ (Exhaustive), $D=4$ (Proposed),$D=8$ (ML), $D=8$ (MAP), $D=8$ (Exhaustive), $D=8$ (Proposed) }
\end{semilogyaxis}
\end{tikzpicture}
\begin{tikzpicture}[scale=0.9]
\begin{axis}[xlabel={$\beta$},title={b)},
ylabel={FER},grid=major,
legend style={at={(0.26,0.985),font=\footnotesize}, legend cell align={left},
anchor=north,legend columns=1}, xmin=0.85,xmax=0.98,ymax=1,ymin=0
]
\addplot[mark=triangle]  coordinates{
  (   0.8500, 0.0258000)
  (   0.8600, 0.0438000)
  (   0.8700, 0.0740000)
  (   0.8800, 0.1350000)
  (   0.8900, 0.2426000)
  (   0.9000, 0.4086000)
  (   0.9100, 0.6044000)
  (   0.9200, 0.7826000)
  (   0.9300, 0.9054000)
  (   0.9400, 0.9686000)
  (   0.9500, 0.9914000)
  (   0.9600, 0.9982000)
  (   0.9700, 0.9996000)
  (   0.9800, 1)
};
\addplot[mark=triangle*]  coordinates{
  (   0.8500, 0.0150000)
  (   0.8600, 0.0268000)
  (   0.8700, 0.0466000)
  (   0.8800, 0.1024000)
  (   0.8900, 0.2036000)
  (   0.9000, 0.3696000)
  (   0.9100, 0.5634000)
  (   0.9200, 0.7594000)
  (   0.9300, 0.8972000)
  (   0.9400, 0.9686000)
  (   0.9500, 0.9900000)
  (   0.9600, 0.9978000)
  (   0.9700, 0.9994000)
  (   0.9800, 1.0000000)
};
\addplot[mark=otimes,blue]  coordinates{
  (   0.8500, 0.0006000)
  (   0.8600, 0.0028000)
  (   0.8700, 0.0100000)
  (   0.8800, 0.0354000)
  (   0.8900, 0.1090000)
  (   0.9000, 0.2520000)
  (   0.9100, 0.4556000)
  (   0.9200, 0.6806000)
  (   0.9300, 0.8478000)
  (   0.9400, 0.9474000)
  (   0.9500, 0.9844000)
  (   0.9600, 0.9960000)
  (   0.9700, 0.9990000)
  (   0.9800, 1)
};
\addplot[mark=+,red]  coordinates{
  (   0.8500, 0.000600)
  (   0.8600, 0.003400)
  (   0.8700, 0.011600)
  (   0.8800, 0.038000)
  (   0.8900, 0.118800)
  (   0.9000, 0.268200)
  (   0.9100, 0.474000)
  (   0.9200, 0.698800)
  (   0.9300, 0.864600)
  (   0.9400, 0.954000)
  (   0.9500, 0.986000)
  (   0.9600, 0.996400)
  (   0.9700, 0.999200)
  (   0.9800, 1)
};
\addplot[mark=square]  coordinates{
  (   0.8500, 0.0096000)
  (   0.8600, 0.0162000)
  (   0.8700, 0.0248000)
  (   0.8800, 0.0414000)
  (   0.8900, 0.0706000)
  (   0.9000, 0.1356000)
  (   0.9100, 0.2412000)
  (   0.9200, 0.4148000)
  (   0.9300, 0.6130000)
  (   0.9400, 0.7894000)
  (   0.9500, 0.9116000)
  (   0.9600, 0.9718000)
  (   0.9700, 0.9916000)
  (   0.9800, 0.9984000)
};
\addplot[mark=square*]  coordinates{
  (   0.8500, 0.0044000)
  (   0.8600, 0.0068000)
  (   0.8700, 0.0114000)
  (   0.8800, 0.0218000)
  (   0.8900, 0.0454000)
  (   0.9000, 0.1000000)
  (   0.9100, 0.2018000)
  (   0.9200, 0.3630000)
  (   0.9300, 0.5822000)
  (   0.9400, 0.7678000)
  (   0.9500, 0.9038000)
  (   0.9600, 0.9674000)
  (   0.9700, 0.9922000)
  (   0.9800, 0.9980000)
};
\addplot[mark=o,blue]  coordinates{
  (   0.8500, 0)
  (   0.8600, 0)
  (   0.8700, 0.0008000)
  (   0.8800, 0.0020000)
  (   0.8900, 0.0102000)
  (   0.9000, 0.0412000)
  (   0.9100, 0.1138000)
  (   0.9200, 0.2602000)
  (   0.9300, 0.4722000)
  (   0.9400, 0.6940000)
  (   0.9500, 0.8616000)
  (   0.9600, 0.9480000)
  (   0.9700, 0.9850000)
  (   0.9800, 0.9964000)
};
\addplot[mark=asterisk,red]  coordinates{
  (   0.8500, 0)
  (   0.8600, 0)
  (   0.8700, 0.0012000)
  (   0.8800, 0.0030000)
  (   0.8900, 0.0130000)
  (   0.9000, 0.0480000)
  (   0.9100, 0.1218000)
  (   0.9200, 0.2764000)
  (   0.9300, 0.4974000)
  (   0.9400, 0.7118000)
  (   0.9500, 0.8718000)
  (   0.9600, 0.9540000)
  (   0.9700, 0.9866000)
  (   0.9800, 0.9970000)
};
\end{axis}
\end{tikzpicture}
\caption{BER and FER for an ATSC 3.0 LDPC code with rate $R=0.1333$. a) BER versus SNR; and b) FER versus $\beta$.
 }\label{figS222}
\end{figure}

In Simulation~3, we investigate the impact of the number of pilots on BER and FER performance with $R=0.2$. The BER and FER are averaged over \(5000\) frames.
Fig.~\ref{figS3} shows the BER and FER performance obtained by two different strategies versus the number of pilots under soft-decision LDPC decoding, where the reconciliation efficiency \(\beta\) is set to \(0.90\) and \(0.92\) for \(D=4\) and \(D=8\), respectively.
It is observed that:
1) both BER and FER improve as the number of pilots increases;
2) the proposed method consistently outperforms the traditional ML-based method and the MAP-based method, and the performance gap narrows as the number of pilots becomes sufficiently large, which aligns with the results shown in Fig.~\ref{figS3}; and
3) the proposed method is relatively insensitive to the number of pilots, achieving good BER and FER performance even with a small number of pilots.

\begin{figure}
\center
\begin{tikzpicture}[scale=0.9]
\begin{semilogyaxis}[xlabel={$M$},title={a)},
ylabel={BER},grid=major,
legend style={at={(0.750,0.99),font=\footnotesize}, legend cell align={left},
anchor=north,legend columns=1},  xmin=100,xmax=6100,ymax=0.25,ymin=0.066
]
\addplot[mark=triangle]  coordinates{
  (  100, 0.216907645061728)
  (  600, 0.161402907407407)
  ( 1100, 0.147002314814815)
  ( 1600, 0.141199444444444)
  ( 2100, 0.137945299382716)
  ( 2600, 0.135202827160494)
  ( 3100, 0.133462984567901)
  ( 3600, 0.133189379629630)
  ( 4100, 0.132522018518518)
  ( 4600, 0.131603444444444)
  ( 5100, 0.131268888888889)
  ( 5600, 0.131055305555556)
  ( 6100, 0.130584185185185)
};
\addplot[mark=triangle*]  coordinates{
  (  100,  0.200817496913580)
  (  600,  0.149305416666667)
  ( 1100,  0.139476756172839)
  ( 1600,  0.134657185185185)
  ( 2100,  0.133195990740741)
  ( 2600,  0.131737120370370)
  ( 3100,  0.130861055555556)
  ( 3600,  0.130722089506173)
  ( 4100,  0.130186391975309)
  ( 4600,  0.129916848765432)
  ( 5100,  0.129743966049383)
  ( 5600,  0.129361271604938)
  ( 6100,  0.129017317901235)
};
\addplot[mark=+,red]  coordinates{
  (  100, 0.134252481481481)
  (  600, 0.128725157407407)
  ( 1100, 0.128333759259259)
  ( 1600, 0.127993157407407)
  ( 2100, 0.127985438271605)
  ( 2600, 0.127691333333333)
  ( 3100, 0.127646743827161)
  ( 3600, 0.127648861111111)
  ( 4100, 0.127733814814815)
  ( 4600, 0.127618617283951)
  ( 5100, 0.127579570987654)
  ( 5600, 0.127585651234568)
  ( 6100, 0.127576345679012)
};
\addplot[mark=square]  coordinates{
  (  100, 0.181929246913580 )
  (  600, 0.104321401234568 )
  ( 1100, 0.0889297067901235)
  ( 1600, 0.0820276141975309)
  ( 2100, 0.0783712962962963)
  ( 2600, 0.0771394135802469)
  ( 3100, 0.0758104660493827)
  ( 3600, 0.0752291080246914)
  ( 4100, 0.0746992253086420)
  ( 4600, 0.0735755956790123)
  ( 5100, 0.0734819907407408)
  ( 5600, 0.0729152716049383)
  ( 6100, 0.0726924012345679)
};
\addplot[mark=square*]  coordinates{
  (  100, 0.157885250000000 )
  (  600, 0.0896321851851852)
  ( 1100, 0.0790647098765432)
  ( 1600, 0.0765749228395062)
  ( 2100, 0.0747471851851852)
  ( 2600, 0.0734193209876543)
  ( 3100, 0.0731347191358025)
  ( 3600, 0.0729791388888889)
  ( 4100, 0.0724308117283951)
  ( 4600, 0.0722130216049383)
  ( 5100, 0.0720377283950617)
  ( 5600, 0.0717448333333333)
  ( 6100, 0.0715990277777778)
};
\addplot[mark=asterisk,red]  coordinates{
  (  100, 0.0790805123456790)
  (  600, 0.0715382685185186)
  ( 1100, 0.0713508148148148)
  ( 1600, 0.0709879351851852)
  ( 2100, 0.0712616234567901)
  ( 2600, 0.0710072037037037)
  ( 3100, 0.0709189228395062)
  ( 3600, 0.0708528734567901)
  ( 4100, 0.0709444691358025)
  ( 4600, 0.0709192962962963)
  ( 5100, 0.0709011203703704)
  ( 5600, 0.0707665061728395)
  ( 6100, 0.0708022222222222)
};
\legend{$D=4$ (ML), $D=4$ (MAP), $D=4$ (Proposed), $D=8$ (ML), $D=8$ (MAP), $D=8$ (Proposed)}
\end{semilogyaxis}
\end{tikzpicture}
\begin{tikzpicture}[scale=0.9]
\begin{axis}[xlabel={$M$},title={b)},
ylabel={FER},grid=major,
legend style={at={(0.26,0.985),font=\footnotesize}, legend cell align={left},
anchor=north,legend columns=1}, xmin=100,xmax=6100,ymax=0.88,ymin=0.37
]
\addplot[mark=triangle]  coordinates{
  (  100, 0.8570)
  (  600, 0.7182)
  ( 1100, 0.6766)
  ( 1600, 0.6588)
  ( 2100, 0.6490)
  ( 2600, 0.6420)
  ( 3100, 0.6346)
  ( 3600, 0.6356)
  ( 4100, 0.6326)
  ( 4600, 0.6314)
  ( 5100, 0.6292)
  ( 5600, 0.6278)
  ( 6100, 0.6286)
};
\addplot[mark=triangle*]  coordinates{
  (  100, 0.8190)
  (  600, 0.6846)
  ( 1100, 0.6546)
  ( 1600, 0.6378)
  ( 2100, 0.6342)
  ( 2600, 0.6298)
  ( 3100, 0.6278)
  ( 3600, 0.6270)
  ( 4100, 0.6252)
  ( 4600, 0.6238)
  ( 5100, 0.6246)
  ( 5600, 0.6228)
  ( 6100, 0.6210)
};
\addplot[mark=+,red]  coordinates{
  (  100,  0.6594 )
  (  600,  0.6262 )
  ( 1100,  0.6230 )
  ( 1600,  0.6206 )
  ( 2100,  0.6204 )
  ( 2600,  0.6208 )
  ( 3100,  0.6188 )
  ( 3600,  0.6192 )
  ( 4100,  0.6192 )
  ( 4600,  0.6190 )
  ( 5100,  0.6174 )
  ( 5600,  0.6178 )
  ( 6100,  0.6194 )
};
\addplot[mark=square]  coordinates{
  (  100, 0.7476)
  (  600, 0.5194)
  ( 1100, 0.4696)
  ( 1600, 0.4482)
  ( 2100, 0.4332)
  ( 2600, 0.4284)
  ( 3100, 0.4230)
  ( 3600, 0.4210)
  ( 4100, 0.4184)
  ( 4600, 0.4142)
  ( 5100, 0.4140)
  ( 5600, 0.4114)
  ( 6100, 0.4100)
};
\addplot[mark=square*]  coordinates{
  (  100,  0.6774 )
  (  600,  0.4706 )
  ( 1100,  0.4364 )
  ( 1600,  0.4246 )
  ( 2100,  0.4176 )
  ( 2600,  0.4130 )
  ( 3100,  0.4126 )
  ( 3600,  0.4120 )
  ( 4100,  0.4086 )
  ( 4600,  0.4070 )
  ( 5100,  0.4062 )
  ( 5600,  0.4056 )
  ( 6100,  0.4048 )
};
\addplot[mark=asterisk,red]  coordinates{
  (  100, 0.4540  )
  (  600, 0.4080  )
  ( 1100, 0.4060  )
  ( 1600, 0.4040  )
  ( 2100, 0.4046  )
  ( 2600, 0.4050  )
  ( 3100, 0.4032  )
  ( 3600, 0.4028  )
  ( 4100, 0.4038  )
  ( 4600, 0.4034  )
  ( 5100, 0.4050  )
  ( 5600, 0.4046  )
  ( 6100, 0.4026  )
};
\end{axis}
\end{tikzpicture}
\caption{BER and FER performance versus the number of pilots for different reconciliation dimensions and efficiencies. a) BER; and b) FER.
 }\label{figS3}
\end{figure}

{In Simulation~4, we evaluate the parameter estimation and  reconciliation performance under short block-length conditions, where the LDPC code defined in the 5G NR standard with base graph~2 (BG2) is adopted. The lifting size is set to \(Z_c=10\) with lifting-set index \(i_{\mathrm{LS}}=2\), resulting in a codeword length of \(N=520\) and a code rate of \(R=0.1923\). The total number of Monte Carlo trials is set to \(J=10^6\), and the reconciliation efficiency \(\beta\) is fixed to \(\beta=0.9\).
Fig.~\ref{figS42}-a) shows the RMSE performance of different strategies for estimating \(t\) as the number of pilot symbols \(M\) varies from 52 to 520, while Fig.~\ref{figS42}-b) shows the BER performance of different strategies versus the number of pilot symbols under soft-decision LDPC decoding.
The results demonstrate that the proposed method still maintains satisfactory performance even when the pilot and data lengths are relatively short, verifying its applicability in fast-fading channel scenarios.
}

\begin{figure}
\center
\begin{tikzpicture}[scale=0.9]
\begin{semilogyaxis}[xlabel={$M$},title={a)},
ylabel={$\mathrm{RMSE}_t$},grid=major,
legend style={at={(0.735,0.98),font=\footnotesize}, legend cell align={left},
anchor=north,legend columns=1},  xmin=52,xmax=520,ymax=0.163,ymin=0.038
]
\addplot[mark=triangle]  coordinates{
  (    52,  0.139346506155254)
  (   104,  0.097404740906948)
  (   156,  0.079442428757953)
  (   208,  0.068496767426570)
  (   260,  0.061246253347290)
  (   312,  0.055858770270835)
  (   364,  0.051691457313680)
  (   416,  0.048307326992697)
  (   468,  0.045521220388860)
  (   520,  0.043169949089570)
};
\addplot[mark=triangle*]  coordinates{
  (    52,  0.135325069139204)
  (   104,  0.096026086891141)
  (   156,  0.078683905441595)
  (   208,  0.068018247352764)
  (   260,  0.060897786327068)
  (   312,  0.055595799010022)
  (   364,  0.051488751579676)
  (   416,  0.048138783308597)
  (   468,  0.045382617068254)
  (   520,  0.043040750934132)
};
\addplot[mark=+,red]  coordinates{
  (    52,  0.126808878001343)
  (   104,  0.089570157650038)
  (   156,  0.073814747343482)
  (   208,  0.064136598150405)
  (   260,  0.057803236153713)
  (   312,  0.052995210346238)
  (   364,  0.049261720772998)
  (   416,  0.046247357396840)
  (   468,  0.043737568589398)
  (   520,  0.041619910018059)
};
\addplot[mark=asterisk,red]  coordinates{
  (    52,  0.123485114229694)
  (   104,  0.087099219987829)
  (   156,  0.071916495449872)
  (   208,  0.062577467453475)
  (   260,  0.056463306617256)
  (   312,  0.051837038756215)
  (   364,  0.048264481910056)
  (   416,  0.045367213138379)
  (   468,  0.042920085232248)
  (   520,  0.040896062992596)
};
\legend{ML, MAP, $D=4$ (Proposed), $D=8$ (Proposed)}
\end{semilogyaxis}
\end{tikzpicture}
\begin{tikzpicture}[scale=0.9]
\begin{semilogyaxis}[xlabel={$M$},title={b)},
ylabel={BER},grid=major,
legend style={at={(0.735,0.98),font=\footnotesize}, legend cell align={left},
anchor=north,legend columns=1},  xmin=52,xmax=520,ymax=0.22,ymin=0.166]
\addplot[mark=triangle]  coordinates{
  (    52,  0.206934723076923)
  (   104,  0.198207455769231)
  (   156,  0.195077803846154)
  (   208,  0.193172163461539)
  (   260,  0.192219942307693)
  (   312,  0.191402328846154)
  (   364,  0.191170557692308)
  (   416,  0.190649586538462)
  (   468,  0.190471069230770)
  (   520,  0.190310290384616)
};
\addplot[mark=triangle*]  coordinates{
  (    52,   0.200501690384616)
  (   104,   0.194007073076923)
  (   156,   0.192038607692308)
  (   208,   0.190874621153846)
  (   260,   0.190431361538462)
  (   312,   0.189899055769231)
  (   364,   0.189906036538462)
  (   416,   0.189546928846154)
  (   468,   0.189521057692308)
  (   520,   0.189451034615385)
};
\addplot[mark=+,red]  coordinates{
  (    52,    0.190706094230770)
  (   104,    0.190091834615385)
  (   156,    0.190013909615385)
  (   208,    0.189770094230770)
  (   260,    0.189692311538462)
  (   312,    0.189505600000000)
  (   364,    0.189679890384616)
  (   416,    0.189383807692308)
  (   468,    0.189501925000000)
  (   520,    0.189448669230770)
};
\addplot[mark=square]  coordinates{
  (    52,   0.188222407692308)
  (   104,   0.178853984615385)
  (   156,   0.175284784615385)
  (   208,   0.173352838461539)
  (   260,   0.172332988461539)
  (   312,   0.171428600000000)
  (   364,   0.171118326923077)
  (   416,   0.170641676923077)
  (   468,   0.170325853846154)
  (   520,   0.170244563461539)
};
\addplot[mark=square*]  coordinates{
  (    52,  0.181123615384616)
  (   104,  0.174403873076923)
  (   156,  0.172171367307693)
  (   208,  0.171052321153846)
  (   260,  0.170625992307693)
  (   312,  0.169969611538462)
  (   364,  0.169921928846154)
  (   416,  0.169565488461539)
  (   468,  0.169415475000000)
  (   520,  0.169401848076923)
};
\addplot[mark=asterisk,red]  coordinates{
  (    52,   0.171601323076923)
  (   104,   0.170535551923077)
  (   156,   0.170163557692308)
  (   208,   0.169923348076923)
  (   260,   0.169892296153846)
  (   312,   0.169505578846154)
  (   364,   0.169624088461539)
  (   416,   0.169431861538462)
  (   468,   0.169293621153846)
  (   520,   0.169442367307692)
};
\legend{$D=4$ (ML), $D=4$ (MAP), $D=4$ (Proposed), $D=8$ (ML),  $D=8$ (MAP), $D=8$ (Proposed)}
\end{semilogyaxis}
\end{tikzpicture}
\caption{{RMSE and BER performance versus the number of pilot symbols under short block-length conditions using a 5G NR LDPC code with \(N=520\) and \(R=0.1923\). a) RMSE for estimating \(t\); b) BER.}} \label{figS42}
\end{figure}

\begin{figure}
  \centering
  \includegraphics{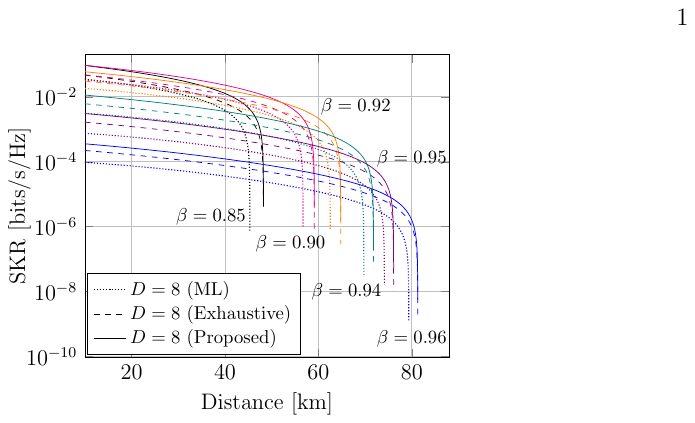}
  \caption{Finite-size SKR of the CV-QKD system employing one-way reverse reconciliation at a repetition rate of $5$~MHz over different transmission distances for an ATSC 3.0 LDPC code with rate $R=0.2$.}\label{FigSKR}
\end{figure}

Finally, we examine the distance dependence of the finite-size SKR in a one-way reverse reconciliation CV-QKD system. The finite-size SKR under homodyne detection is mathematically formulated as:
\begin{align}
\mathrm{SKR} = \frac{N(1 - \mathrm{FER})}{M+N} \bigl( \beta I(A;B) - {S_{\epsilon_{PE}}(B;E)} - \Lambda(N) \bigr),
\end{align}
where \(I(A;B)\) denotes the mutual information between Alice and Bob,
{\(S_{\epsilon_{PE}}(B;E)\) denotes the Holevo information corresponding to the parameter-estimation security level \(\epsilon_{PE}\), and \(\Lambda(N)\simeq 7\sqrt{\frac{\log_2(2/\bar{\epsilon})}{N}}\) is related to the security of privacy amplification with the smoothing parameter \(\bar{\epsilon}=10^{-10}\) \cite{PhysRevA.81.062343}.
}
The channel transmittance is given as:
\begin{align}
T(d) = 10^{-\alpha d / 10},
\end{align}
where \(\alpha = 0.2\,\mathrm{dB/km}\) is the standard loss coefficient and \(d\) denotes the transmission distance in kilometers. The excess channel noise is set to \(\xi = 0.01\) in shot-noise units.
Bob's homodyne detector efficiency is set at \(\eta = 0.606\) with an added electronic noise of \(v_{el} = 0.041\). To simplify the selection of Alice's modulation variance and ensure that the SNR remains constant across the entire distance range, we set \cite{WOS:000432546600001}:
\begin{align}
V_A(\beta, \zeta) = \kappa(\beta) \big(1 + \chi_{\mathrm{tot}}(\zeta)\big),
\end{align}
where \(\kappa(\beta)\) is defined in (\ref{eq-kappa}), and \(\chi_{\mathrm{tot}}(\zeta) = \chi_{\mathrm{line}}(\zeta) + \frac{\chi_{\mathrm{hom}}}{T(\zeta)}\) denotes the total noise referred to the input of the quantum channel with
\begin{align}
\chi_{\mathrm{line}}(\zeta) &= \frac{1}{T(\zeta)} + \xi - 1, \\
\chi_{\mathrm{hom}} &= \frac{1 + v_{el}}{\eta} - 1.
\end{align}
Fig.~\ref{FigSKR} presents the finite-size SKR performance of a 5~MHz CV-QKD system employing one-way reverse reconciliation, with the number of pilots set to \(M=600\) for both the proposed method and the ML method, and \(M=64800\) for the exhaustive ML method. It is observed that: 1) the ML method yields the lowest SKR performance and the shortest transmission distance due to its poor channel parameter estimation; 2) the exhaustive ML method achieves the same transmission distance as the proposed method, but with a much lower SKR, as it requires a large number of pilots, resulting in a significant loss of symbol efficiency; and 3) the proposed method demonstrates superior overall performance, which is attributed to its improved FER characteristics (see Fig.~\ref{figS2}-b), enabled by joint parameter estimation and multidimensional reconciliation.

\section{Conclusion}
In this paper, we introduced a unified message-passing framework for CV-QKD that jointly performs quantum channel parameter estimation and information reconciliation. By embedding the estimation task within a Bayesian decoding process using the EM algorithm, our method eliminates the dependence on large pilot overhead and the drawbacks of conventional maximum likelihood estimators. The integration of a generalized factor graph enables simultaneous LDPC decoding and parameter learning, mitigating error propagation between disjoint phases. Additionally, the proposed hybrid multidimensional rotation scheme removes the need for norm feedback, further reducing classical communication overhead. Overall, our joint design provides a practical and efficient solution to high-accuracy parameter estimation and reconciliation in CV-QKD, representing a significant step toward more symbol-efficient and robust quantum key distribution systems.


%

\bibliography{apssamp}

\end{document}